\documentclass[12pt,letterpaper]{article}

\usepackage{ccn}
\usepackage{pslatex}
\usepackage{apacite}
\usepackage[labelfont=bf]{caption}
\usepackage{url}

\usepackage{graphicx}
\graphicspath{ {figs/} }

\usepackage{changepage}
\usepackage{fancyhdr}

\title{\huge Deep Learning for Cognitive Neuroscience}
\author{{}
\AND{\large \bf Katherine R. Storrs\textsuperscript{1} and Nikolaus Kriegeskorte\textsuperscript{2}}\\
1. Abteilung Allgemeine Psychologie, Justus-Liebig University Giessen, Germany\\
2. Mortimer B. Zuckerman Mind Brain Behavior Institute,\\Department of Psychology, Department of Neuroscience,\\Department of Electrical Engineering (Affiliated member), Columbia University, USA}

\begin{document}

\maketitle

\begin{adjustwidth*}{1cm}{1cm}
\rule{0.9\textwidth}{1pt}
\begin{quote}
\linespread{1.3}
\selectfont{} % required to refresh the line spacing command

Neural network models can now recognise images, understand text, translate languages, and play many human games at human or superhuman levels. These systems are highly abstracted, but are inspired by biological brains and use only biologically plausible computations. In the coming years, neural networks are likely to become less reliant on learning from massive labelled datasets, and more robust and generalisable in their task performance. From their successes and failures, we can learn about the computational requirements of the different tasks at which brains excel. Deep learning also provides the tools for testing cognitive theories. In order to test a theory, we need to realise the proposed information-processing system at scale, so as to be able to assess its feasibility and emergent behaviours. Deep learning allows us to scale up from principles and circuit models to end-to-end trainable models capable of performing complex tasks. There are many levels at which cognitive neuroscientists can use deep learning in their work, from inspiring theories to serving as full computational models. Ongoing advances in deep learning bring us closer to understanding how cognition and perception may be implemented in the brain -- the grand challenge at the core of cognitive neuroscience.

\end{quote}
\rule{0.9\textwidth}{1pt}
\end{adjustwidth*}

\linespread{1.5}
\selectfont{} % required to refresh the line spacing command

\section{Introduction: neuro-inspired AI and AI-inspired neuroscience}

To explain how brains reason, remember, perceive, and act, theories must bridge from biology to behaviour. Rigorous explanation requires models that use neurobiologically plausible components to implement cognitive processes (Poeppel, 2012; Kriegeskorte \& Douglas, 2018; Kriegeskorte \& Mok, 2017). Artificial neural networks, composed of simplified simulated neurons, have long promised such models (McCulloch \& Pitts, 1943; Rumelhart \& McClelland, 1986). Thanks to computational and methodological advances, neural networks now outperform all other engineering solutions to pattern-recognition problems (e.g. Russakovsky et al., 2015), as well as many cognitive challenges such as language translation (e.g. Wu et al., 2016), visual reasoning (e.g. Santoro et al., 2017) and game-playing (e.g. Mnih et al., 2015; Jaderberg et al., 2018).

Ideas from psychology and neuroscience have inspired engineers and underlie many features in modern networks (Kriegeskorte, 2015; Hassabis et al., 2017). Networks used for visual tasks process images hierarchically with spatially restricted receptive fields as does mammalian visual cortex (LeCun et al., 1998; Russakovsky et al., 2015). "Attention networks" dynamically select subparts of their inputs to which they sequentially devote processing resources (e.g. Xu et al., 2015). Research on the complementary roles of the hippocampus and neocortex in learning (Kumaran et al., 2016; McLelland, McNaughton \& O'Reilly, 1995) has inspired networks with something akin to episodic memory, which bootstrap their learning via "experience replay" (e.g. Mnih et al., 2015).

Reciprocally, cognitive neuroscientists have been inspired and informed by ideas and results from deep learning. Neural networks provide proofs of concept which help address difficult questions, such as the degree to which visual object recognition requires recurrent processing (Riesenhuber \& Poggio, 1999; O'Reilly et al., 2013; Spoerer, McClure \& Kriegeskorte, 2017). Machine learning offers new perspectives, encouraging us to think about cortical and subcortical specialisation in terms of the learning objectives present in different brain regions and the prior world knowledge, which may be ingrained in regionally specific cytoarchitectures (Marblestone \& Kording, 2016). The humbling experience of trying to teach machines to see and think has impressed upon neuroscientists just how computationally difficult these achievements are. Engineering has also demonstrated how much can be done with relatively simple computational components (e.g. the "GIST" model of Oliva \& Torralba, 2001).

Neural network models excel at pattern recognition and pattern generation tasks, whether the patterns are static or dynamic. They do not yet capture the human ability to make comprehensive inferences from tiny details, learn new concepts from single experiences, and generalise successfully to entirely novel domains (Lake et al., 2017, Kriegeskorte 2018). More abstract cognitive computational models, based on symbolic representations and probabilistic inference, currently come closer to matching these amazing human cognitive feats, although they fall short of the human brain in terms of computational efficiency (Gershman, Horvitz \& Tenenbaum, 2015) and are harder to relate to neurobiology. Understanding human cognition and its implementation in the brain will require both cognitive-level computational models and neural network models. Here we focus on the latter, and invite cognitive neuroscientists to incorporate these models into their research.

\section{What are deep neural networks?}

Models of brain computation range from biologically detailed simulations of circuits of spiking model neurons to highly abstract cognitive models. The term "neural network model" usually refers to models at an intermediate level of abstraction. Neural network models consist of interconnected units, each computing a weighted sum of its inputs and passing it through a nonlinear activation function (Figure 1a). The resulting scalar output can be thought of as the firing rate of an idealised neuron (called "unit"), with an instantaneous response and no adaptation or refractory period. Networks of such units can implement arbitrarily complex functions between inputs (e.g. photographs) and outputs (e.g. the name of the main object in each photograph) as compositions of linear-nonlinear subfunctions.

\begin{figure}[ht!]
\begin{center}
\includegraphics[width=0.90\textwidth]{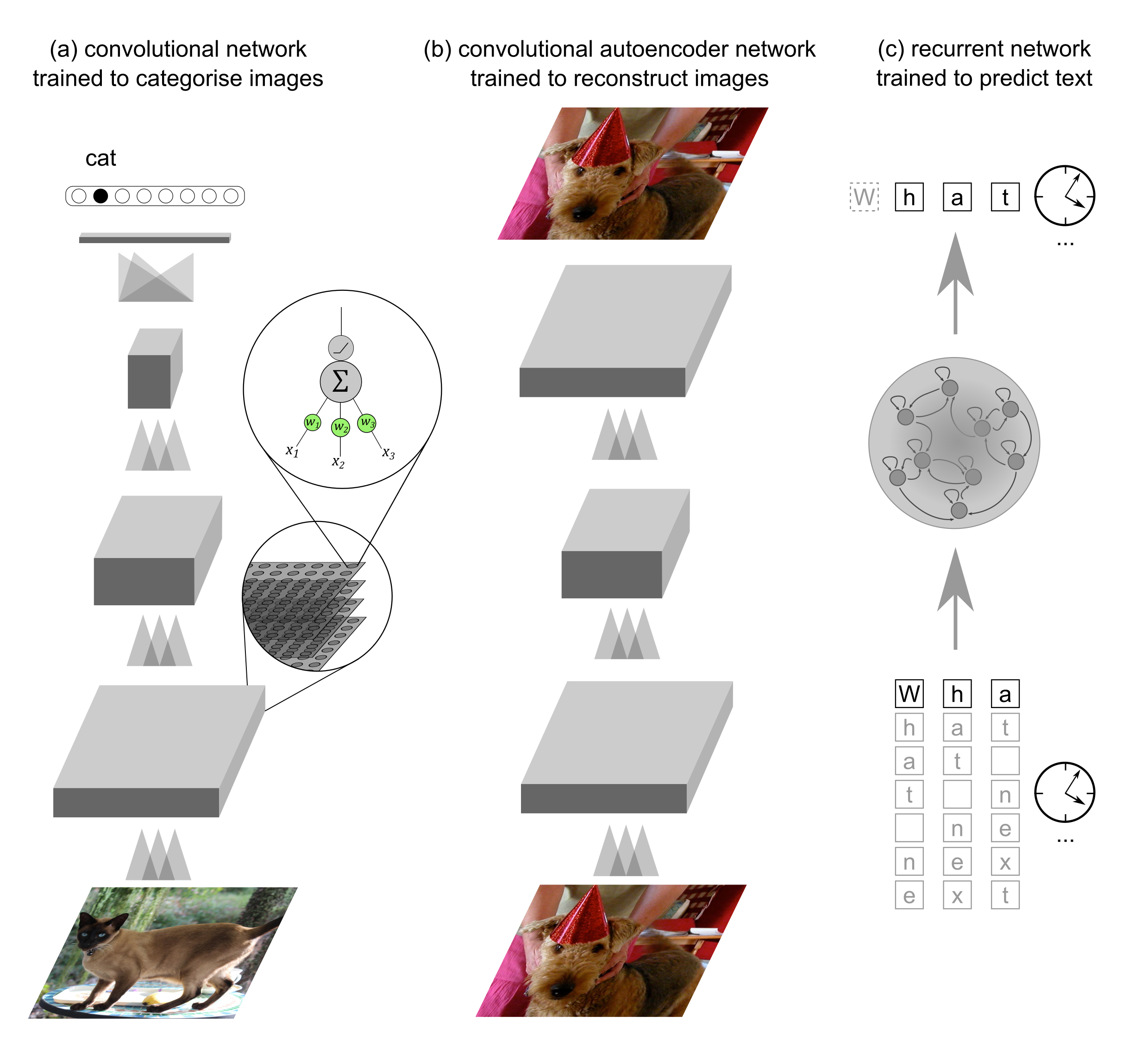}
\end{center}
\linespread{1}
\caption{Diagrams of some of the species of neural network. \textbf{(a)} A convolutional neural network with four hidden layers (three convolutional and one fully connected) which is able, after supervised training on labelled images, to categorise objects in novel photographs it has not been trained with. The convolutional layers consist of multiple "feature maps," each of which contains units with spatially restricted receptive fields centred on different locations of the input, but sharing the same connection weights. This allows each map to represent the presence or absence of a particular feature at each location in the image. \textbf{(b)} A convolutional autoencoder with three convolutional hidden layers, which is able, after unsupervised training, to compress novel images into a more concise format within the middle "bottleneck" layer, and approximately reconstruct them thereafter. \textbf{(c)} A recurrent neural network which takes as input text, one character at a time, and is able, after unsupervised training, to predict the next character in sentences it has never seen before.} 
\end{figure}
\linespread{1.5}

\subsection{Learned connection weights: the flexible knowledge of a network}

Like biological neural networks, the models don’t come into the world knowing everything about their environment, or how to perform the tasks that will be required of them. Instead, they learn. The knowledge inside a network is held by the weights associated with connections between units, loosely analogous to synaptic weights in a brain. Weights are usually set initially to small random numbers, and incrementally updated via a learning algorithm.

Learning algorithms fall into three broad types: supervised, unsupervised, and reinforcement. In supervised learning, the learning algorithm will adjust the weights so as to bring the outputs for a large set of input patterns closer to pre-specified desired outputs. A cost captures how much the current outputs deviate from the desired outputs. Weights are adjusted in proportion to how strongly their adjustment reduces the cost. This requires computing the derivative of the cost function with respect to each weight. An efficient algorithm for computing these derivatives is backpropagation. The derivative of the cost for a weight tells us in which direction and how much to tweak each connection weight in order to bring the output for a training input closer to the desired output. Many recent engineering achievements, for example in machine vision and language translation, owe their success to supervised learning on massive ground-truth-labelled datasets (Russakovsky et al., 2015; Wu et al., 2016).

In unsupervised learning, training data are provided without any labels. The network learns to model the statistics of its input data. An autoencoder, for example, learns to compress its input pattern within a lower-dimensional "code" layer. The network maps the input pattern to the lower-dimensional code (encoder component) and then back to the full input pattern (decoder component). Autoencoders learn to encode and decode (i.e. reconstruct) each input pattern (e.g. Figure 1b). They can also learn to remove noise or predict future states of a dynamic input. Unsupervised learning signals are extremely rich (an image autoencoder derives a training signal for every pixel of its reconstruction attempt), enabling the network to better exploit the information in its experiential data. The network attempts to learn all regularities in its data, not only those relevant for a particular task.

Finally, in reinforcement learning (Sutton \& Barto, 1998), the network outputs an action (e.g. a movement in a simulated or real environment), and learns to model the expected cumulative reward associated with possible actions. Certain events in the environment are defined as rewarding, and their occurrence drives learning. Reinforcement learning can be combined with deep learning, so as to enable the value function to be represented by a neural network (e.g. Deep Q-Learning; Mnih et al., 2015). When rewards are few and far between in the environment, reinforcement learning is difficult, because it provides fewer direct constraints for adjusting the weights than unsupervised or supervised learning. Rooted in biology and psychology, however, reinforcement learning has high ecological plausibility. It has recently also brought significant advances in engineering, illustrated, for example, by its success in the domain of video game playing (Mnih et al., 2015; Jaderberg et al., 2018).

Networks will overfit their training data to some extent and are therefore always tested on an independent test set of novel inputs (e.g. images) and outputs (e.g. category labels). Unlike in biological brains, learning (the adjustment of connection weights) and perception (the processing sweep of a new input through a network with fixed connection weights) are typically separate processes, and learning generally takes place in a distinct initial phase.

\subsection{Architecture: the fixed structure of a network}

Infinitely many network architectures can be created by linking up units in different configurations. In feedforward architectures, units form a single processing hierarchy, with no unit connected to itself or any previous unit. Feedforward networks have no temporal dynamics beyond the feedforward sweep, and compute a static mapping from inputs to outputs. A network that contains one or more loops in its directed connection graph, such as top-down feedback, is recurrent, and its internal state will evolve over time in discrete steps. Recurrent networks are often used to process time-series data such as video or text, with a new frame or character being fed into the network at each time step (Figure 1c).

A feedforward network with only a single "hidden" layer of units between its input data and output responses is known as a \textit{shallow} network. Shallow feedforward networks can already approximate any continuous function, with increasing precision as the number of units increases (Hornik, 1991). However, adding more intermediate processing layers can allow the network to express more complex functions with the same number of units by letting later units reuse and recombine features calculated by previous units. A network is called \textit{deep} when it has more than one hidden layer intervening between input and output units. In modern computer vision, deep neural networks capable of near-human image classification performance typically contain ten or more hidden layers, and over a million individual units (e.g. Simonyan \& Zisserman, 2015; He et al., 2016).

Specialised architectures and unit types can exploit prior knowledge about the domain or task. For example, deep \textit{convolutional} neural networks (CNNs) are often used for visual tasks (Figure 1a). CNNs are loosely inspired by mammalian visual cortex and use units with spatially restricted receptive fields, and shared weights (LeCun et al., 1998). Weight-sharing means that multiple units within each layer use the same template of connection weights, applying this template at different locations on the input. This reduces the number of parameters the network must learn, and captures the prior belief that the same feature (e.g. a vertical edge, or the wing of a bird) may appear anywhere within an image. Many recurrent networks use "long short-term memory" (LSTM) units (Hochreiter \& Schmidhuber, 1997), which can dynamically retain recent information for as long as it is useful in the current task. Other custom units perform local response normalisation or max-pooling over their inputs, inspired by the neuroscientific idea of canonical cortical computations (Carandini \& Heeger, 2012; Riesenhuber \& Poggio, 1999).

\section{Using neural networks as models of cognition and perception}

Cognitive neuroscientists can engage with neural networks at different levels, requiring varying degrees of technical know-how and resource commitment. At one end of the spectrum, we can read and think about networks, using their successes, failures, and principles to seed theories and hypotheses. We can consider task-performing models as proofs of principle for computational mechanisms of perception and cognition. For example, the extent to which visual object recognition can be achieved by feedforward systems has been a matter of historical contention (e.g. Riesenhuber \& Poggio, 1999). Modern CNNs demonstrate that object naming and segmentation in natural scenes can be achieved, to a large extent, without lateral or feedback connections. At the same time, recurrent connectivity can substantially improve the recognition performance of neural networks under challenging circumstances (O'Reilly et al., 2013; Spoerer, McClure \& Kriegeskorte, 2017; Nayebi et al., 2018).

At a deeper level of engagement, we can begin to bring DNN models into our empirical work by taking pre-trained networks built by other researchers (many of which are available online), and comparing them to human and nonhuman subjects in terms of their behaviour and internal representations. To disentangle the contributions of architecture and training to a model’s ability to explain brain representations and behaviour, we can retrain or fine-tune pre-built architectures on stimuli and tasks that relate to our particular hypotheses.

Finally, we can develop new models, exploring different training data, learning objectives and architectures. This lets us test how different aspects of the environment, learning process, and neural structure may affect cognitive function. Designing complex neural network models and performing the behavioural and neural experiments to evaluate them requires a wide range of expertise that is not easy to integrate in a single lab. We will therefore need to develop new forms of collaboration and sharing across labs. Some labs may choose to focus more on building models, others on testing shared models with brain and behavioural data, or on developing tasks designed to highlight and quantify remaining shortcomings of existing models.

\subsection{Testing neural network models with behavioural and brain-activity data}

At the behavioural level, a model should be able to perform the task of interest at a similar level to a human or animal subject. However, a model merely being able to perform well in a task in which humans perform well (e.g. generating a verbal description of an image; Xu et al., 2015) does not provide strong evidence that the model performs similar computations or arrives at the output via similar internal representations as humans do. Good models will be able to predict detailed patterns of behavioural variation across different instances of the task.

We can also compare each model's internal representations to human perceptual judgements. The similarity of stimuli or conditions in the model's internal activation patterns should predict perceived similarity to humans (Kubilius, Bracci \& Op de Beeck, 2016). Stimuli that elicit identical responses within the model should appear identical to humans (Wallis et al., 2017). As stimuli are degraded or distorted, the model's performance should decline in a quantitatively similar way to human performance (contrary to this, Geirhos et al. (2018) found that CNNs were far less robust to image noise than humans). Models should be able to predict patterns of confusions and errors, ideally at the single-stimulus level. For example, in one study CNN classifications accurately predicted object-matching confusions in monkeys and humans at the object-category level, but failed to predict which specific images were confused (Rajalingham et al., 2018).

At the level of the internal representation, a model should go through the same sequences of representational transformations across space (brain regions) and time (sequence of processing). Comparing the internal representations between model and brain is complicated by the fact that we may not know the detailed spatial and temporal correspondence mapping between model activity patterns and brain activity patterns. Diedrichsen (2018) introduces the framework of representational models, which can be used to test neural network models, by comparing their internal representations with brain-activity measurements. Briefly, encoding models predict each measured response channel as a linear combination of the units of the neural network (Kay et al., 2008; Dumoulin et al., 2008; Mitchell et al., 2008; Naselaris \& Gallant, 2011). Representational similarity analysis (Kriegeskorte et al., 2008; Nili et al., 2014; Kriegeskorte \& Diedrichsen, 2016) and pattern component modelling (Diedrichsen et al., 2011) use a stimulus-by-stimulus matrix of dissimilarities or similarities to characterise the representational space and compare model layers to brain regions. These approaches have complementary pros and cons. For example, encoding models lend themselves to analysing responses in each voxel separately and mapping these out over the cortex (G{\"u}{\c c}l{\"u} \& van Gerven 2015; Eickenberg et al., 2017; Wen et al., 2017); representational similarity analysis and pattern component modeling obviate the need for fitting thousands of parameters of a linear encoding model, while still allowing the flexibility of estimating weights to model the relative prevalence of different features in a brain representation (e.g. Khaligh-Razavi \& Kriegeskorte, 2014; Khaligh-Razavi, Henriksson, Kay \& Kriegeskorte, 2017). Encoding models, representational similarity analysis, and pattern component modelling are best thought of as part of a toolbox of representational modelling techniques that can be combined as appropriate to the goals of a study (Diedrichsen \& Kriegeskorte, 2016).

Khaligh-Razavi \& Kriegeskorte (2014) found that the later layers of a categorisation-trained CNN outperformed any of 27 shallow computer vision models in predicting the representation of natural object images in inferior temporal cortex. Categorisation-trained CNNs also best predict visual cortical activity whether measured via electrophysiology (Cadieu et al., 2014), fMRI (G{\"u}{\c c}l{\"u} \& van Gerven 2015; Eickenberg et al., 2017; Wen et al., 2017), MEG (Cichy et al., 2016; Cichy et al., 2017) or EEG (Greene \& Hansen, 2018), and a DNN trained on speech and music best predicted auditory cortical activity (Kell et al., 2018). Variants of this approach which focus on decoding or reconstructing stimuli from neural activity likewise find that CNNs provide excellent feature spaces for decoding activity in visual areas (Wen et al., 2017; Horiwaka \& Kamitani, 2017).

Whatever method is used to evaluate a DNN model, it is crucial to include comparisons to control models, and/or to test multiple DNNs that instantiate alternative theories of computation or learning. Even trivial models of stimulus processing explain small but significant variance in sensory brain regions (e.g. the pixel-based control models in Khaligh-Razavi \& Kriegeskorte, 2014), as do randomly-weighted untrained DNNs (e.g. Cichy et al., 2016).

\subsection{Interpreting neural network models: doing cognitive neuroscience \textit{in silico}}

Once we have a DNN that can perform some task and explain brain and behavioural data (up to the limit determined by noise and intersubject variability), we have reached an important milestone. However, our job is not done. Because the model has been trained on the task and has many parameters, we may not understand its computational mechanism. We therefore need to analyse the representations and dynamics in the model. Like brains, DNNs can be studied at many different levels of resolution and abstraction. Unlike brains, they permit perfect access to the entire system. We can stop and restart a network, have it re-learn under different environments or task demands, gather data from it continuously without fatigue or damage, lesion and reinstate any combination of its components, use stimulus optimisation techniques to see what features it has learned, and even analytically prove some of its properties\footnote{For example, the mathematics of group theory have been used to explain why successive layers of deep feedforward networks learn increasingly complex features (Paul \& Venkatasubramanian, 2014).}.

Because of our unfettered access, DNNs are far more amenable to "electrophysiological" methods than real brains are. To analyse the feature preferences learned by individual units in a network, experimenters can present thousands or millions of stimuli and record each unit's activation (e.g. Yosinski et al., 2015), or complementarily, occlude different regions of an image stimulus to find which features are most crucial for each unit's activation (Zhou et al., 2014). Figure 2c shows how the activation of a single LSTM unit inside a recurrent text-prediction network trained on movie reviews changes as the passage the network is predicting unfolds (Radford, Jozefowicz \& Sutskever, 2018). In this striking example, the visualised unit appears to have learned to represent whether the movie review is expressing a positive or negative sentiment.

DNNs are also amenable to novel interrogation techniques that could never be applied to a brain. Because neural network models are differentiable, it is possible to use gradient-descent optimisation to create stimuli that elicit specific patterns of network activation, including generating optimal stimuli for individual units (e.g. Yosinski et al., 2015). Several techniques exist for doing this, as beautifully summarised and illustrated by Olah (2017). Figure 2a shows a noise image which has been iteratively optimised to strongly activate each of four different layers in two category-trained CNNs using Google's "DeepDream" algorithm (Mordvintsev, Olah \& Tyka, 2015). The flow of information can be traced through networks to reveal where or when in the stimulus evidence was drawn in order to support a certain output decision (again beautifully illustrated by Olah (2018)), or to which part of the stimulus a network with dynamic "attention" is currently devoting its resources (Figure 2b).

Direct optimisation methods have revealed that the hierarchy of features learned by deep visual networks bears a striking similarity to that in the mammalian ventral stream (see e.g. Yamins \& DiCarlo, 2016). They also allow us to explore how feature preferences develop over the course of training, suggesting tantalising similarities with human perceptual learning (Wenliang \& Seitz, 2018). Task-performing networks may help settle longstanding debates in cognitive neuroscience such as how sparse or distributed neural codes are likely to be (Agrawal, Girschick \& Malik, 2014; Zhuang, Wang, Yamins \& Hu, 2017; Morcos, Barrett, Rabinowitz \& Botvinick, 2018).

Network representations can also be summarised at higher levels of abstraction. Traditional functional localisation methods can identify units or layers that are preferentially selective for particular classes of stimuli, with initially counter-intuitive results such as the emergence of object-selective units in a visual network trained only to classify scenes (Zhou et al., 2014). Representational similarity analysis (RSA) can help show how a network transforms stimulus information across the layers (e.g. Khaligh-Razavi \& Kriegeskorte, 2014) or time-steps (e.g. Cichy et al., 2016) of its processing.

The accessibility of neural network models raises new possibilities for efficient model search and falsification that could substantially change how we design experiments. Currently experimental conditions are usually chosen "by hand" to differentiate among hypotheses suggested by verbal theories. In the future, conditions could be optimised algorithmically to best differentiate between explicit computational models (e.g. Wang \& Simoncelli, 2008).

\begin{figure}[ht!]
\begin{center}
\includegraphics[width=0.90\textwidth]{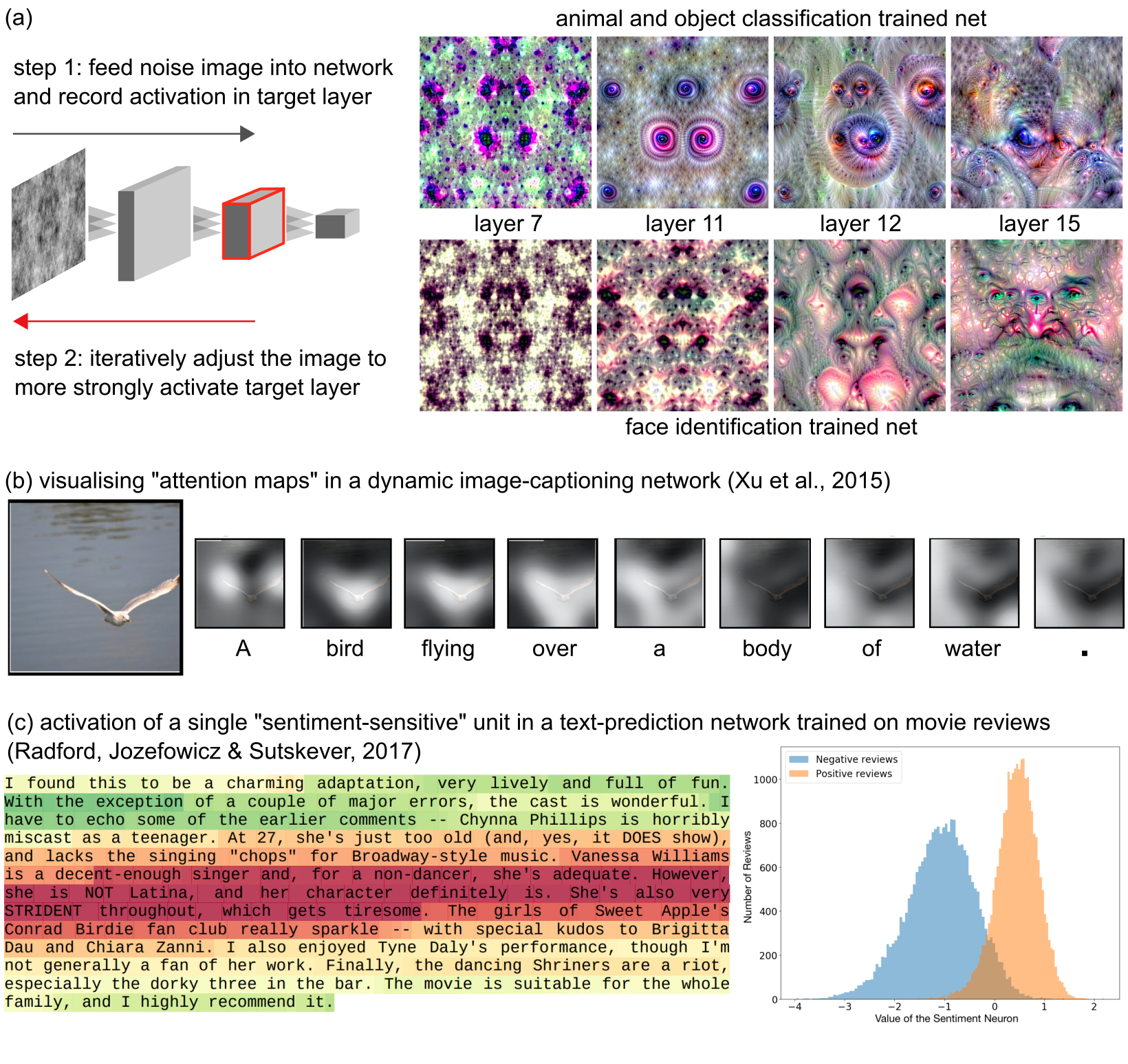}
\end{center}
\linespread{1}
\caption{The relative transparency of deep neural networks. \textbf{(a)} The backpropagation algorithm used to train networks can also be used to visualise features in trained networks. Here, a noise image has been iteratively optimised to increase how strongly it activates units in each of four layers in a 16-layer DCNN, which has been trained either to categorise objects and animals (top row) or identify faces (bottom row). Training has a striking impact on the features learned. \textbf{(b)} "Attention networks" have a spatial attention mask (a multiplicative weighting of the input) which they learn to allocate in ways that improve performance at their training task. Here, the white patches show which regions of this image were attended to at each time step by a recurrent network trained to generate text captions of images, outputting one word per time step (reproduced with permission from Xu et al., 2015). \textbf{(c)} Visualisation of the output activation of a single LSTM unit in a recurrent network trained to predict text one character at a time, for an example passage the network has never seen before. High activation is indicated by green, and low by red. Responses of this unit, when analysed across a large sample of reviews, predicted whether a review was positive or negative in sentiment (reproduced with permission from Radford, Jozefowicz \& Sutskever, 2018).} 
\end{figure}
\linespread{1.5}

\section{Neural networks as models of diverse cognitive functions}

Visual object and face recognition have received a great deal of attention from engineers and cognitive neuroscientists because DCNNs have been able to reach near-human-level performance using static stimuli and a single feedforward sweep of processing (He et al., 2015; Taigman et al., 2014). However, the deep learning revolution has touched almost all domains in cognition and perception, including our higher-level abilities such as language and reasoning.

\subsection{Language in neural networks}

Language-processing DNNs are widely used in automatic translation, for example of websites and online search results. Such high-level tasks tend to be achieved by more complex compositional network architectures than we have yet encountered. The state-of-the-art Google Neural Machine Translation system (Wu et al., 2016) consists of two eight-layer recurrent neural networks, one which encodes the text from its original language to a latent representation within the network, and another which takes the latent representation and decodes it into the target language, using "attentional" weights to change which parts of the encoded representation it processes at each time step. The entire dynamic process is learned through backpropagation by training on tens of millions of human-translated sentences, including the form of the latent representation and the allocation of attention during decoding (Wu et al., 2016). A similar approach has been used to generate verbal descriptions of images (Xu et al., 2016). There, an image-processing convolutional network is used to transform the image into a latent representation of high-level image features, and a recurrent network is trained to output a sequence of words based on those features, using spatial attention to modulate which parts of the image most guide word choice at each time step (Figure 2b).

Although these networks are engineering solutions, it is striking how the composition of multiple pattern-recognition DNNs, coupled with attentional guidance (a psychologically inspired feature), can enable such uniquely human skills as language translation and image description. Cognitive neuroscientists are just beginning to bring such principles into neural network models of human cognition and language. Devereux, Clarke and Tyler (2018) used a modular visual-semantic model to predict fMRI activation patterns as human observers named objects in images. In the model, images were first processed by a feedforward image CNN trained on object classification, and the penultimate-layer features from the CNN were then fed into a single-layer recurrent network, which was trained to encode the presence or absence of a large number of semantic properties of the pictured objects (\textit{"is a fruit"}, \textit{"grows on trees"}, etc.). Similarity of representations within the CNN better predicted similarity in human visual cortex, while similarity within the recurrent semantic network better predicted similarity in human perirhinal cortex, an area associated with semantic processing.

\subsection{Reasoning in neural networks}

Artificial general intelligence may not yet be in sight, but some deep networks display impressively flexible reasoning. Relation Networks (Santoro et al., 2017) achieved superhuman performance on the CLEVR benchmark, a notoriously difficult task involving reasoning about the relations among objects in images. Example questions include \textit{"How many objects have the same shape as the green object?"} and \textit{"Are there any rubber things that have the same size as the yellow metallic cylinder?"} (Santoro et al., 2017). A feedforward image CNN extracts high-level visual features from the image, and a recurrent language network encodes the verbal question. The Relation Network then considers all possible pairs of "objects" in the image (operationalised as pairs of spatial locations), along with the question, and evaluates the likelihood of each possible response. Relation Networks achieved a 95\% accuracy rate on the CLEVR task; human performance is around 93\% (Santoro et al., 2017). The image, language, and relation-reasoning sub-networks form and end-to-end differentiable system and are simply trained together using backpropagation.

Compositional network architectures have proved an important engineering trick for achieving intelligent behaviour in DNNs. Recent work by Yang, Song, Newsome \& Wang (2017) begins to unpick how task requirements might determine neural architectures. A recurrent network was trained simultaneously to perform twenty very simple tasks from human and animal cognitive research, such as speeded classification and delayed match-to-sample. The authors were then able to analyse which units were involved in which tasks. They found that the network had learned a compositional representation of the twenty tasks, in that clusters of units specialised in the components of the tasks (e.g. \textit{"remember the direction of the cue"} or \textit{"respond in the direction opposite to the cue"}), and any given task was performed by a coalition of clusters, according to the composition of its sub-tasks.

\subsection{Embodied neural networks}

DNNs are not necessarily disembodied. For example, a recurrent network trained to simulate the muscle movements made by monkeys reaching towards targets in an experiment learned internal representations that predicted the firing patterns of motor neurons recorded while the monkeys were reaching (Sussilo, Churchland, Kaufman \& Shenoy, 2015). Neural networks trained to navigate in a simulated environment developed units with similar tuning properties to place, grid, border, and head-direction cells found in rat entorhinal cortex (Cueva \& Wei, 2018; Banino et al., 2018). Embodied models, whether with physical or virtual bodies, may be able to learn more efficiently than disembodied ones by using active learning methods to seek out informative training data (e.g. Haber, Mrowca, Fei-Fei \& Yamins, 2018).

\section{Current challenges for deep neural networks in cognitive neuroscience}

Cognitive neuroscientists have much work to do. Any particular current instance of a deep neural network, presented in the machine learning literature for a particular practical application, will almost certainly not prove to be a fully satisfying model of how similar functions are performed by human brains. We must seek out ways to make cognition in DNNs more flexible and generalisable, learning more ecologically plausible, and turn the predictive power of neural network models into theoretical understanding. Happily, many of these goals align at least in part with those of industrial AI researchers.

\subsection{Modelling more robust and flexible cognition and perception}

Although DNNs are architecturally deep, they can seem conceptually shallow. Their intelligence is more fragile than that of humans, apparently failing to develop the causal and conceptual understanding that allows us to generalise performance across tasks and environments (Marcus, 2017; Lake, Ullman, Tenenbaum \& Gershman, 2017). Visual object-recognition DNNs can be tricked into drastic misclassifications using minute image perturbations imperceptible to humans (Szegedy et al., 2013). Reading comprehension networks able to answer questions based on passages of text can be similarly tricked by small changes to the text which do not distract human readers (Jia \& Liang, 2017). Standard DNNs are poor at estimating the confidence of their decisions, and may confidently return seemingly nonsense classifications when shown stimuli with different statistics than those in their training dataset (Nguyen, Yosinski \& Clune, 2015; Gharamanhi, 2015). Reinforcement-learning networks can outcompete humans at playing arcade video games (Mnih et al., 2015), but appear to depend on features which humans would consider superficial or irrelevant -- changing the shape or position of the player's paddle by just a few pixels can disastrously undermine the network's performance (Kansky et al., 2017).

It is worth bearing in mind that current DNNs are still very limited in computational scale compared to biological brains. A modern DNN might contain hundreds of thousands or millions of units, on the same order as the number of cortical neurons in a handful of fMRI voxels\footnote{The successful image recognition network AlexNet has around 658,000 units (Khrizhevsky, 2012), and there is a density of around 630,000 neurons per 3mm isotropic voxel in the human cortex (\url{https://cfn.upenn.edu/aguirre/wiki/public:neurons_in_a_voxel}).}. Importantly, the units in neural network models cannot be equated to biological neurons, which are more complex in structure and dynamics, and thus potentially more computationally powerful. As a result, DNNs likely fall short by a larger factor than counting units and neurons would suggest. The numbers of units and neurons simply cannot meaningfully be compared.

To compound their disadvantage in scale, DNNs so far have tended to live exclusively within a single modality and task. At best this makes them models of specific sensory processing areas, but even these areas in human brains have extensive interaction with other modalities, executive control, and action. The examples described in the section "Reasoning in neural networks" above hint at some of the possible solutions, involving compositional architectures (Santoro et al., 2017; Xu et al., 2015), as well as hybrid solutions which combine neural networks with symbolic reasoning models (e.g. Battaglia et al., 2016; Evans \& Grefenstette, 2018).

Cognitive neuroscience is well placed to devise benchmark tasks that capture important hallmarks of human cognition. Many existing experimental paradigms may be ideal, for example "violation of expectation" tasks used to measure the development of intuitive physics concepts in children can be applied to assess models' reasoning about physical interactions (Piloto, 2018). Lake and Baroni (2017) and Marcus (2017) suggest a number of other tasks to evaluate the profundity and generality of machine cognition. Formalising these as benchmarks would provide concrete goals for the modelling community and drive progress (Kriegeskorte \& Douglas 2018), as object recognition benchmarks have done in computer vision (Russakovsky et al., 2015).

\subsection{Building networks that learn in more ecologically and biologically plausible ways}

Both supervised and unsupervised learning depend on the backpropagation algorithm, which is traditionally considered biologically implausible (Glaser, Benjamin, Farhoodi \& Kording, 2018). Recent theoretical work suggests that deep learning with an algorithm similar to backpropagation might be biologically feasible (Lillicrap et al., 2016; Guerguiev, Lillicrap, and Richards, 2017; Kording \& K{\"o}nig 2001; Scellier \& Bengio, 2016; Hinton and McClelland 1988). However, supervised training on millions of labelled stimuli is still an ecologically unrealistic requirement. Animal brains are capable learning far more data-efficiently, often without explicit supervision, and sometimes even from single examples.

Unsupervised, weakly-supervised and one-shot learning are current focuses of the machine learning community (Hassabis et al., 2017), and we can expect advances in these areas over the next few years. Cognitive neuroscientists may be able to draw from brain theory and data for creative and biologically feasible solutions. One influential idea within neuroscience is that \textit{prediction} may be fundamentally important to how brains learn and perceive (Hawkins \& Blakeslee, 2007). Predictive coding theories, in particular, propose that, during perception, neural activity primarily encodes the differences between predicted and actual sensory information (Srinivasan, Laughlin \& Dubs, 1982). Explicit computational models of predictive coding have previously been formulated at the neuronal-circuit level (Rao \& Ballard, 1999), and have received some support from electrophysiological (Rao \& Ballard, 1999) and fMRI data (Muckli et al., 2015). In machine learning, prediction offers a rich unsupervised training signal requiring no additional reward or label information. When implemented in a modern deep learning framework, a predictive coding network was capable of predicting future frames in video of natural environments (Lotter, Kreiman \& Cox, 2017; 2018). Furthermore, the network spontaneously discovered, in its deeper layers, higher-level properties of the objects depicted in videos, such as facial identity and pose (Lotter, Kreiman \& Cox, 2017), and its individual units reproduced certain temporal dynamics of primate visual neurons (Lotter, Kreiman \& Cox, 2018). "Curiosity-based" learning is another tantalising method motivated by animal learning, in which networks embodied in simulated environments actively seek out the most informative parts of those environments during learning (Haber, Mrowca, Fei-Fei \& Yamins, 2018).

\subsection{Illuminating the black box: from prediction to explanation}

A criticism levelled at neural networks as cognitive models is that success at predicting brain and behavioural data, which DNNs have unarguably enjoyed (Khaligh-Razavi et al. 2014; Cadieu et al. 2014; G{\"u}{\c c}l{\"u} \& van Gerven 2015; Cichy et al., 2016;  Eickenberg et al., 2017; Kubilius et al., 2016; Wen et al., 2017), does not constitute success at explaining or understanding the brain or mind (Kay, 2017). If a model exhaustively predicted neural and behavioural data with generalisation to new instances of the task, we might be convinced that it is an accurate model of a certain cognitive function. But unless we can express more concisely how it performs that function, we are unlikely to be fully satisfied as scientists.

We have described some techniques of \textit{in silico} electrophysiology and visualisation of internal representations for illuminating the black box above. An even more satisfying approach is to go from a neural network model to a concise mathematical description. Gon{\c c}alves and Welchmann (2017) trained a convolutional neural network to judge the relative depth of objects in scenes from pairs of stereo images, requiring it to solve the classic "correspondence problem" (which points in the two retinal images correspond to the same point on an external object?). By tracing the strength and sign of connection weights in the trained network, they discovered a surprising computational strategy: rather than using positive weights to find positive matches between points in the two images, the network primarily used negative weights to suppress incorrect matches. The authors distilled this strategy into a succinct mathematical image-filtering model. The model displayed several unusual and previously unexplained phenomena in human depth perception, and substantially changes our understanding of the correspondency problem, which has stood as a computational puzzle for at least 100 years (Gon{\c c}alves \& Welchmann, 2017). Deep neural network models do not replace intuitive explanations, verbal theories, and concise mathematical descriptions. Instead they make complex hypotheses testable and help us bridge the levels of description between verbally communicable theory and neural implementation.

\section{Conclusion}

Deep neural networks are currently at the cutting edge of artificial intelligence, and are the most powerful predictive models yet discovered for many aspects of behaviour (Gon{\c c}alves \& Welchmann, 2017; Kubilius, Bracci \& Op de Beeck, 2016; Wallis et al., 2017), single-neuron activity (e.g. Cadieu et al., 2014), and large-scale cortical activity (Khaligh-Razavi \& Kriegeskorte, 2014; Cichy et al., 2016; Cichy et al., 2017; Wen et al., 2017; G{\"u}{\c c}l{\"u} \& van Gerven 2015; Eickenberg et al., 2017; Greene \& Hansen, 2018; Kell et al., 2018; Horiwaka \& Kamitani, 2017).  With biologically plausible components and structures, they are the closest we have yet come to explicit end-to-end models of how perception and cognition might be performed in brains.

Deep learning, therefore, is relevant to everyone interested in how perception and cognition arise from neural activity. There are many levels at which cognitive neuroscientists can use deep learning in their work, from considering the modelling literature in theoretical discussions, through testing models build by others, and on to building new models based on cognitive and neuroscientific theories. Conversely, the cognitive neuroscience community may help drive engineering progress, bringing DNNs that learn in more ecologically feasible ways, learn more profound conceptual representations, and display more robust and generalisable task performance. Machine learning of computations capable of real-world tasks in biologically plausible systems will play a major role in understanding how intelligent behaviour arises from brains.

\clearpage

\section{Appendix: resources for getting started with deep learning}
\subsection{Articles}

There are many excellent review articles on cognitive neuroscientific applications of deep neural network models. For a recent overview of the complementary exchanges between neuroscience and artificial intelligence, see Hassabis et al. (2017). For a more extensive treatment of neural network architectures, principles, and their application as models of brain function, see Kriegeskorte (2015) and Kietzmann, McClure \& Kriegeskorte (2017). For reviews of neural network models in sensory and systems neuroscience, see Yamins \& DiCarlo (2016) and Glaser, Benjamin, Farhoodi \& Kording (2018).

\subsection{Tutorials, courses and books}

For accessible introductions to machine learning and deep neural networks, the free online courses currently offered by Geoff Hinton (\url{https://www.coursera.org/learn/neural-networks}) and Andrew Ng (\url{https://www.coursera.org/learn/machine-learning}) are invaluable, as is the online book by Michael Nielsen (\url{http://neuralnetworksanddeeplearning.com/}). Deep Learning (Goodfellow, Bengio \& Courville, 2016) is a comprehensive book written by some of the leaders in the field.

\subsection{Software}

Software frameworks to implement deep neural network models are rapidly evolving. At the time of writing, the leading frameworks are free and Python-based: Tensorflow (developed by Google), PyTorch (developed by Facebook) and Caffe (developed by Berkeley University). Keras and Sonnet are higher-level packages which work atop Tensorflow to provide readier access to many advanced features. MATLAB (MathWorks, Inc.) is less widely supported or used by machine learning teams in industry, but is worth mentioning due to its popularity in psychology labs. The Neural Network Toolbox in versions of MATLAB from 2017 onward allows all core deep learning operations such as defining and training feedforward and recurrent models on CPUs or GPUs, loading pre-trained models, importing models defined in other frameworks (currently Keras and Caffe are supported), and visualising units within trained networks.

\clearpage

\section{References}
\setlength{\leftskip}{4em}
\setlength{\parindent}{-4em}

Agrawal, P., Girshick, R., \& Malik, J. (2014). Analyzing the performance of multilayer neural networks for object recognition. In \textit{Proceedings of the European Conference on Computer Vision} (pp. 329-344).

Banino, A., Barry, C., Uria, B., Blundell, C., Lillicrap, T., Mirowski, P., ... \& Wayne, G. (2018). Vector-based navigation using grid-like representations in artificial agents. \textit{Nature, 557}(7705), 429.

Battaglia, P., Pascanu, R., Lai, M., \& Rezende, D. J. (2016). Interaction networks for learning about objects, relations and physics. In \textit{Advances in Neural Information Processing Systems} (pp. 4502-4510).

Cadieu, C. F., Hong, H., Yamins, D. L., Pinto, N., Ardila, D., Solomon, E. A., ... \& DiCarlo, J. J. (2014). Deep neural networks rival the representation of primate IT cortex for core visual object recognition. \textit{PLoS Computational Biology, 10}(12), e1003963.

Carandini, M., \& Heeger, D. J. (2012). Normalization as a canonical neural computation. \textit{Nature Reviews Neuroscience, 13}(1), 51.

Cichy, R. M., Khosla, A., Pantazis, D., Torralba, A., \& Oliva, A. (2016). Comparison of deep neural networks to spatio-temporal cortical dynamics of human visual object recognition reveals hierarchical correspondence. \textit{Scientific Reports, 6}, 27755.

Cichy, R. M., Khosla, A., Pantazis, D., \& Oliva, A. (2017). Dynamics of scene representations in the human brain revealed by magnetoencephalography and deep neural networks. \textit{NeuroImage, 153}, 346-358.

Cueva, C. J., \& Wei, X. X. (2018). Emergence of grid-like representations by training recurrent neural networks to perform spatial localization. \textit{arXiv} preprint. \url{arXiv:1803.07770}

Devereux, B. J., Clarke, A. D., \& Tyler, L. K. (2018). Integrated deep visual and semantic attractor neural networks predict fMRI pattern-information along the ventral object processing pathway. \textit{Scientific Reports, 8}, 1.

Diedrichsen, J. (2018). Representational models and the feature fallacy. To appear in M. Gazzaniga (Ed.), \textit{The Cognitive Neurosciences} (6th Edition). Boston: MIT Press.

Diedrichsen, J., \& Kriegeskorte, N. (2017). Representational models: A common framework for understanding encoding, pattern-component, and representational-similarity analysis. \textit{PLoS Computational Biology, 13}(4), e1005508.

Diedrichsen, J., Ridgway, G. R., Friston, K. J., \& Wiestler, T. (2011). Comparing the similarity and spatial structure of neural representations: a pattern-component model. \textit{NeuroImage, 55}(4), 1665-1678.

Dumoulin, S. O., \& Wandell, B. A. (2008). Population receptive field estimates in human visual cortex. \textit{NeuroImage, 39}(2), 647-660.

Eickenberg, M., Gramfort, A., Varoquaux, G., \& Thirion, B. (2017). Seeing it all: Convolutional network layers map the function of the human visual system. \textit{NeuroImage, 152}, 184-194.

Evans, R., \& Grefenstette, E. (2018). Learning explanatory rules from noisy data. \textit{Journal of Artificial Intelligence Research, 61}, 1-64.

Geirhos, R., Temme, C. R. M., Rauber, J., Schuett, H. H., Bethge, M., \& Wichmann, F. A. (2018). Generalisation in humans and deep neural networks. \textit{arXiv} preprint. \url{arXiv:1808.08750}

Gershman, S. J., Horvitz, E. J., \& Tenenbaum, J. B. (2015). Computational rationality: A converging paradigm for intelligence in brains, minds, and machines. \textit{Science, 349}(6245), 273-278.

Ghahramani, Z. (2015). Probabilistic machine learning and artificial intelligence. \textit{Nature, 521}(7553), 452.

Glaser, J. I., Benjamin, A. S., Farhoodi, R., \& Kording, K. P. (2018). The roles of supervised machine learning in systems neuroscience. \textit{arXiv} preprint. \url{arXiv:1805.08239}

Gon{\c c}alves, N. R., \& Welchman, A. E. (2017). "What not" detectors help the brain see in depth. \textit{Current Biology, 27}(10), 1403-1412.

Goodfellow, I., Bengio, Y., Courville, A., \& Bengio, Y. (2016). \textit{Deep learning (Vol. 1)}. Cambridge: MIT press.

Greene, M. R., \& Hansen, B. C. (2018). Shared spatiotemporal category representations in biological and artificial deep neural networks. \textit{PLoS Computational Biology, 14}(7), e1006327.

G{\"u}{\c c}l{\"u}, U., \& van Gerven, M. A. (2015). Deep neural networks reveal a gradient in the complexity of neural representations across the ventral stream. \textit{Journal of Neuroscience, 35}(27), 10005-10014.

Guerguiev, J., Lillicrap, T. P., \& Richards, B. A. (2017). Towards deep learning with segregated dendrites. \textit{eLife, 6}, e22901.

Haber, N., Mrowca, D., Fei-Fei, L., \& Yamins, D. L. (2018). Emergence of structured behaviors from curiosity-based intrinsic motivation. \textit{arXiv} preprint. \url{arXiv:1802.07461}

Hassabis, D., Kumaran, D., Summerfield, C., \& Botvinick, M. (2017). Neuroscience-inspired artificial intelligence. \textit{Neuron, 95}(2), 245-258.

Hawkins, J., \& Blakeslee, S. (2007).\textit{ On intelligence: How a new understanding of the brain will lead to the creation of truly intelligent machines.} Macmillan.

He, K., Zhang, X., Ren, S., \& Sun, J. (2016). Deep residual learning for image recognition. In \textit{Proceedings of the IEEE Conference on Computer Vision and Pattern Recognition} (pp. 770-778).

Hinton, G. E., \& McClelland, J. L. (1988). Learning representations by recirculation. In \textit{Advances in Neural Information Processing System}s (pp. 358-366).

Horikawa, T., \& Kamitani, Y. (2017). Generic decoding of seen and imagined objects using hierarchical visual features. \textit{Nature Communications, 8}, 15037.

Hornik, K. (1991). Approximation capabilities of multilayer feedforward networks. \textit{Neural Networks, 4}(2), 251-257.

Huys, Q.J., Maia, T.V., \& Frank, M.J. (2016). Computational psychiatry as a bridge from neuroscience to clinical applications. \textit{Nature Neuroscience, 19}(3), 404.

Jaderberg, M., Czarnecki, W. M., Dunning, I., Marris, L., Lever, G., Castaneda, A. G., ... \& Sonnerat, N. (2018). Human-level performance in first-person multiplayer games with population-based deep reinforcement learning. \textit{arXiv} preprint. \url{arXiv:1807.01281}

Jia, R., \& Liang, P. (2017). Adversarial Examples for Evaluating Reading Comprehension Systems. \textit{arXiv} preprint. \url{arXiv:1707.07328}

Kansky, K., Silver, T., M{\'e}ly, D. A., Eldawy, M., L{\'a}zaro-Gredilla, M., Lou, X., ... \& George, D. (2017). Schema networks: Zero-shot transfer with a generative causal model of intuitive physics. \textit{arXiv} preprint. \url{arXiv:1706.04317}

Kay, K. N., Naselaris, T., Prenger, R. J., \& Gallant, J. L. (2008). Identifying natural images from human brain activity. \textit{Nature, 452}(7185), 352.

Kay, K. N. (2017). Principles for models of neural information processing. \textit{NeuroImage, 180}, 101-109.

Kell, A. J., Yamins, D. L., Shook, E. N., Norman-Haignere, S. V., \& McDermott, J. H. (2018). A task-optimized neural network replicates human auditory behavior, predicts brain responses, and reveals a cortical processing hierarchy. \textit{Neuron, 98}(3), 630-644.

Kording, K. P., \& K{\"o}nig, P. (2001). Supervised and unsupervised learning with two sites of synaptic integration. \textit{Journal of Computational Neuroscience, 11}(3), 207-215.

Khaligh-Razavi, S. M., \& Kriegeskorte, N. (2014). Deep supervised, but not unsupervised, models may explain IT cortical representation. \textit{PLoS Computational Biology, 10}(11), e1003915.

Khaligh-Razavi, S. M., Henriksson, L., Kay, K., \& Kriegeskorte, N. (2017). Fixed versus mixed RSA: Explaining visual representations by fixed and mixed feature sets from shallow and deep computational models. \textit{Journal of Mathematical Psychology, 76}, 184-197.

Kietzmann, T. C., McClure, P., \& Kriegeskorte, N. (2018). Deep neural networks in computational neuroscience. In \textit{Oxford Research Encyclopedia of Neuroscience}, Oxford University Press.

Kriegeskorte, N., Mur, M., \& Bandettini, P. A. (2008). Representational similarity analysis-connecting the branches of systems neuroscience. \textit{Frontiers in Systems Neuroscience, 2}, 4.

Kriegeskorte (2015) Deep neural networks: A new framework for modeling biological vision and brain information processing. \textit{Annual Review of Vision Science, 1}, 417-446.

Kriegeskorte, N., \& Diedrichsen, J. (2016). Inferring brain-computational mechanisms with models of activity measurements. \textit{Philosophical Transactions of the Royal Society B, 371}(1705), 20160278.

Kriegeskorte, N., \& Douglas, P. K. (2018). Cognitive computational neuroscience. \textit{Nature Neuroscience, 21}, 1148–1160.

Kriegeskorte, N., \& Mok, R. M. (2017). Building machines that adapt and compute like brains. \textit{Behavioral and Brain Sciences, 40}.

Krizhevsky, A., Sutskever, I., \& Hinton, G. E. (2012). Imagenet classification with deep convolutional neural networks. In \textit{Advances in Neural Information Processing Systems} (pp. 1097-1105).

Kubilius, J., Bracci, S., \& Op de Beeck, H. P. (2016). Deep neural networks as a computational model for human shape sensitivity. \textit{PLoS Computational Biology, 12}(4), e1004896.

Kumaran, D., Hassabis, D., \& McClelland, J. L. (2016). What learning systems do intelligent agents need? Complementary learning systems theory updated. \textit{Trends in Cognitive Sciences, 20}(7), 512-534.

Lake, B. M., Ullman, T. D., Tenenbaum, J. B., \& Gershman, S. J. (2017). Building machines that learn and think like people. \textit{Behavioral and Brain Sciences, 40}.

Lake, B. M., \& Baroni, M. (2017). Still not systematic after all these years: On the compositional skills of sequence-to-sequence recurrent networks. \textit{arXiv} preprint \url{arXiv:1711.00350}

LeCun, Y., Bottou, L., Bengio, Y., \& Haffner, P. (1998). Gradient-based learning applied to document recognition. In \textit{Proceedings of the IEEE, 86}(11), 2278–2324. 

Lillicrap, T. P., Cownden, D., Tweed, D. B., \& Akerman, C. J. (2016). Random synaptic feedback weights support error backpropagation for deep learning. \textit{Nature Communications, 7}, 13276.

Lotter, W., Kreiman, G., \& Cox, D. (2016). Deep predictive coding networks for video prediction and unsupervised learning. \textit{arXiv} preprint. \url{arXiv:1605.08104}

Lotter, W., Kreiman, G., \& Cox, D. (2018). A neural network trained to predict future video frames mimics critical properties of biological neuronal responses and perception. \textit{arXiv} preprint. \url{arXiv:1805.10734}

Marblestone, A. H., Wayne, G., \& Kording, K. P. (2016). Toward an integration of deep learning and neuroscience. \textit{Frontiers in Computational Neuroscience, 10}, 94.

Marcus, G. (2018). Deep learning: A critical appraisal. \textit{arXiv} preprint. \textit{arXiv:1801.00631}

McCulloch, W. \& Pitts, W. (1943). A logical calculus of the ideas immanent in nervous activity. \textit{Bulletin of Mathematical Biophysics, 5}, 115–133.

McClelland, J. L., McNaughton, B. L., \& O'Reilly, R. C. (1995). Why there are complementary learning systems in the hippocampus and neocortex: Insights from the successes and failures of connectionist models of learning and memory. \textit{Psychological Review, 102}, 419–457.

Mitchell, T. M., Shinkareva, S. V., Carlson, A., Chang, K. M., Malave, V. L., Mason, R. A., \& Just, M. A. (2008). Predicting human brain activity associated with the meanings of nouns. \textit{Science, 320}(5880), 1191-1195.

Mnih, V., Kavukcuoglu, K., Silver, D., Rusu, A. A., Veness, J., Bellemare, M. G., ... \& Petersen, S. (2015). Human-level control through deep reinforcement learning. \textit{Nature, 518}(7540), 529.

Morcos, A. S., Barrett, D. G., Rabinowitz, N. C., \& Botvinick, M. (2018). On the importance of single directions for generalization. \textit{arXiv} preprint. \url{arXiv:1803.06959}

Mordvintsev, Olah \& Tyka (2015) Inceptionism: Going deeper into neural networks. Google technical blog post, retrieved from: \url{https://ai.googleblog.com/2015/06/inceptionism-going-deeper-into-neural.html}

Muckli, L., De Martino, F., Vizioli, L., Petro, L. S., Smith, F. W., Ugurbil, K., ... \& Yacoub, E. (2015). Contextual feedback to superficial layers of V1. \textit{Current Biology, 25}(20), 2690-2695.

Naselaris, T., Kay, K. N., Nishimoto, S., \& Gallant, J. L. (2011). Encoding and decoding in fMRI. \textit{NeuroImage, 56}(2), 400-410.

Nayebi, A., Bear, D., Kubilius, J., Kar, K., Ganguli, S., Sussillo, D., ... \& Yamins, D. L. (2018). Task-driven convolutional recurrent models of the visual system. \textit{arXiv} preprint. \url{arXiv:1807.00053}

Nguyen, A., Yosinski, J., \& Clune, J. (2015). Deep neural networks are easily fooled: High confidence predictions for unrecognizable images. In \textit{Proceedings of the IEEE Conference on Computer Vision and Pattern Recognition} (pp. 427-436).

Nili, H., Wingfield, C., Walther, A., Su, L., Marslen-Wilson, W., \& Kriegeskorte, N. (2014). A toolbox for representational similarity analysis. \textit{PLoS Computational Biology, 10}(4), e1003553.

Olah (2017). Feature visualization. Retrieved from: \url{https://distill.pub/2017/feature-visualization/}

Olah (2018). The building blocks of interpretability. Retrieved from: \url{https://distill.pub/2018/building-blocks/}

Oliva, A., \& Torralba, A. (2001). Modeling the shape of the scene: A holistic representation of the spatial envelope. \textit{International Journal of Computer Vision, 42}(3), 145-175.

O'Reilly, R. C., Wyatte, D., Herd, S., Mingus, B., \& Jilk, D. J. (2013). Recurrent processing during object recognition. \textit{Frontiers in Psychology, 4}, 124.

Paul, A., \& Venkatasubramanian, S. (2014). Why does deep learning work?: A perspective from group theory. \textit{arXiv} preprint. \url{arXiv:1412.6621}

Piloto, L., Weinstein, A., Ahuja, A., Mirza, M., Wayne, G., Amos, D., ... \& Botvinick, M. (2018). Probing physics knowledge using tools from developmental psychology. \textit{arXiv} preprint. \url{arXiv:1804.01128}

Poeppel, D. (2012). The maps problem and the mapping problem: two challenges for a cognitive neuroscience of speech and language. \textit{Cognitive Neuropsychology, 29}(1-2), 34-55.

Radford, A., Jozefowicz, R., \& Sutskever, I. (2017). Learning to generate reviews and discovering sentiment. \textit{arXiv} preprint. \url{arXiv:1704.01444}

Rajalingham, R., Issa, E. B., Bashivan, P., Kar, K., Schmidt, K., \& DiCarlo, J. J. (2018). Large-scale, high-resolution comparison of the core visual object recognition behavior of humans, monkeys, and state-of-the-art deep artificial neural networks. \textit{Journal of Neuroscience, 38}(33), 7255-7269.

Rao, R. P., \& Ballard, D. H. (1999). Predictive coding in the visual cortex: a functional interpretation of some extra-classical receptive-field effects. \textit{Nature Neuroscience, 2}(1), 79.

Riesenhuber, M., \& Poggio, T. (1999). Hierarchical models of object recognition in cortex. \textit{Nature Neuroscience, 2}(11), 1019.

Rumelhart, D.E. \& McClelland, J.L. (1986). \textit{Parallel distributed processing: explorations in the microstructure of cognition}. MIT Press:Cambridge, MA.

Russakovsky, O., Deng, J., Su, H., Krause, J., Satheesh, S., Ma, S., ... \& Berg, A. C. (2015). Imagenet large scale visual recognition challenge. \textit{International Journal of Computer Vision, 115}(3), 211-252.

Santoro, A., Raposo, D., Barrett, D. G., Malinowski, M., Pascanu, R., Battaglia, P., \& Lillicrap, T. (2017). A simple neural network module for relational reasoning. In \textit{Advances in Neural Information Processing Systems} (pp. 4967-4976).

Scellier, B., \& Bengio, Y. (2016). Equilibrium propagation: Bridging the gap between energy-based models and backpropagation. \textit{arXiv} preprint. \url{arXiv:1602.05179}

Hochreiter, S. \& Schmidhuber, J. (1997). Long short-term memory. \textit{Neural Computation, 9}(8), 1735–1780. 

Simonyan, K., \& Zisserman, A. (2015). Very deep convolutional networks for large-scale image recognition. \textit{arXiv} preprint. \url{arXiv:1409.1556}

Spoerer, C. J., McClure, P., \& Kriegeskorte, N. (2017). Recurrent convolutional neural networks: A better model of biological object recognition. \textit{Frontiers in Psychology, 8}, 1551.

Srinivasan, M. V., Laughlin, S. B., \& Dubs, A. (1982). Predictive coding: A fresh view of inhibition in the retina. \textit{Proceedings of the Royal Society of London B, 216}(1205), 427-459.

Sussillo, D., Churchland, M. M., Kaufman, M. T., \& Shenoy, K. V. (2015). A neural network that finds a naturalistic solution for the production of muscle activity. \textit{Nature Neuroscience, 18}(7), 1025.

Sutton, R. S., \& Barto, A. G. (1998). \textit{Introduction to reinforcement learning (Vol. 135)}. Cambridge: MIT press.

Szegedy, C., Zaremba, W., Sutskever, I., Bruna, J., Erhan, D., Goodfellow, I., \& Fergus, R. (2013). Intriguing properties of neural networks. \textit{arXiv} preprint. \url{arXiv:1312.6199}

Taigman, Y., Yang, M., Ranzato, M. A., \& Wolf, L. (2014). Deepface: Closing the gap to human-level performance in face verification. In \textit{Proceedings of the IEEE Conference on Computer Vision and Pattern Recognition} (pp. 1701-1708).

Wallis, T. S., Funke, C. M., Ecker, A. S., Gatys, L. A., Wichmann, F. A., \& Bethge, M. (2017). A parametric texture model based on deep convolutional features closely matches texture appearance for humans. \textit{Journal of Vision, 17}(12), 5-5.

Wang, Z., \& Simoncelli, E. P. (2008). Maximum differentiation (MAD) competition: A methodology for comparing computational models of perceptual quantities. \textit{Journal of Vision, 8}(12), 8-8.

Wen, H., Shi, J., Zhang, Y., Lu, K. H., Cao, J., \& Liu, Z. (2017). Neural encoding and decoding with deep learning for dynamic natural vision. \textit{Cerebral Cortex}, 1-25.

Wenliang, L., \& Seitz, A. R. (2018). Deep neural networks for modeling visual perceptual learning. \textit{Journal of Neuroscience, 38}(27), 6028-6044.

Wu, Y., Schuster, M., Chen, Z., Le, Q. V., Norouzi, M., Macherey, W., ... \& Klingner, J. (2016). Google's neural machine translation system: Bridging the gap between human and machine translation. \textit{arXiv} preprint. \url{arXiv:1609.08144}

Xu, K., Ba, J., Kiros, R., Cho, K., Courville, A., Salakhudinov, R., ... \& Bengio, Y. (2015). Show, attend and tell: Neural image caption generation with visual attention. In \textit{International Conference on Machine Learning} (pp. 2048-2057).

Yamins, D. L. \& DiCarlo, J. J. (2016) Using goal-driven deep learning models to understand sensory cortex. \textit{Nature Neuroscience, 19}(3), 356.

Yang, G. R., Song, H. F., Newsome, W. T., \& Wang, X. J. (2017). Clustering and compositionality of task representations in a neural network trained to perform many cognitive tasks. \textit{bioRxiv} preprint. \url{doi:10.1101/183632}

Yosinski, J., Clune, J., Nguyen, A., Fuchs, T., \& Lipson, H. (2015). Understanding neural networks through deep visualization. \textit{arXiv} preprint. \url{arXiv:1506.06579}

Zhou, B., Khosla, A., Lapedriza, A., Oliva, A., \& Torralba, A. (2014). Object detectors emerge in deep scene CNNs. \textit{arXiv} preprint. \url{arXiv:1412.6856}

Zhuang, C., Wang, Y., Yamins, D. L., \& Hu, X. (2017). Deep learning predicts correlation between a functional signature of higher visual areas and sparse firing of neurons. \textit{Frontiers in Computational Neuroscience, 11}, 100.

\section{Acknowledgments}

This project has received funding from the European Union’s Horizon 2020 Programme for Research and Innovation under the Specific Grant Agreement No. 720270 and 785907 (Human Brain Project SGA1 and SGA2). 

\end{document}